\documentclass[aps,preprint,tightenlines,floatfix]{revtex4}
\usepackage{epsfig}
\usepackage{amsmath}

\preprint{
\vbox{
\hbox{JLAB-THY-02-34}
}}

\begin{document}

\title{Quark-Hadron Duality in Electron-Pion Scattering}
\author{W. Melnitchouk}
\affiliation{Jefferson Lab,
	12000 Jefferson Avenue,
	Newport News, VA 23606 \\}

\begin{abstract}
We explore the relationship between exclusive and inclusive
electromagnetic scattering from the pion, focusing on the transition
region at intermediate $Q^2$.
Combining Drell-Yan data on the leading twist quark distribution in the
pion with a model for the resonance region at large $x$, we calculate
QCD moments of the pion structure function over a range of $Q^2$, and
quantify the role of higher twist corrections.
Using a parameterization of the pion elastic form factor and
phenomenological models for the $\pi\to\rho$ transition form factor,
we further test the extent to which local duality may be valid for the
pion.
\end{abstract}

\maketitle

\newpage

\section{Introduction}

The nature of the transition between quark and hadron degrees of freedom
in QCD is one of the most fundamental problems in strong interaction
physics.
This transition has been extensively explored within nonperturbative
models of QCD, which, while retaining some of the apposite features of
QCD, make simplifying assumptions that allow approximate solutions to be
found \cite{MODELS}.
Considerable progress has also been made recently in calculating
hadronic properties directly from QCD via lattice gauge theory, and much
is anticipated from this approach in the near future with significant
advances in computing power available \cite{LATT}.
It is clear, however, that while a quantitative description of hadronic
structure from first principles in QCD is still some time away,
phenomenological input will remain crucial in guiding our understanding
for the foreseeable future.

Of course, assuming QCD can ultimately describe the physics of hadrons,
the transition from quarks and gluons to hadrons can be considered
trivial in principle from the point of view of quark--hadron duality.
So long as one has access to a complete set of states, it is immaterial
whether one calculates physical quantities in terms of elementary quark
or effective hadron degrees of freedom.
In practice, however, truncations are unavoidable, and it is precisely
the consequences of working with incomplete or truncated basis states
that allows one to expose the interesting dynamics that drives the
quark--hadron transition.

The duality between quarks and hadrons reveals itself in spectacular
fashion through the phenomenon of Bloom-Gilman duality in inclusive
lepton--nucleon scattering, $e N \to e X$.
Here the inclusive $F_2$ structure function of the nucleon measured in
the region dominated by low-lying nucleon resonances is observed to
follow a global scaling curve describing the high energy data, to which
the resonance structure function averages \cite{BG,NICU}.
The equivalence of the averaged resonance and scaling structure
functions in addition appears to hold for each prominent resonance
region separately, suggesting that the resonance--scaling duality also
exists to some extent locally.

The correspondence between exclusive and inclusive observables in
electroproduction was studied even before the advent of QCD
\cite{DY,WEST,BJORKEN,GBB,LANDSHOFF,EZAWA}.
Within QCD, the appearance of duality for the moments of structure
functions can be related through the operator product expansion (OPE)
to the size of high twist corrections to the scaling structure function
\cite{RUJ}, which reflect the importance of long-range multi-parton
correlations in the hadron \cite{MANKIEWICZ}.
The apparent early onset of Bloom-Gilman duality for the proton
structure function seen in recent Jefferson Lab experiments \cite{NICU}
indicates the dominance of single-quark scattering to rather low
momentum transfer \cite{CM}.
It is not {\em a priori} clear, however, whether this is due to an
overall suppression of coherent effects in inclusive scattering, or
because of fortuitous cancellations of possibly large corrections.
Indeed, there are some indications from models of QCD that the workings
of duality may be rather different in the neutron than in the proton
\cite{IJMV,CI}, or for spin-independent and spin-dependent structure
functions.

{}From another direction, one knows from the large $N_c$ limit of QCD
\cite{LARGENC} that duality is an inevitable consequence of quark
confinement; in the mesonic sector one can prove (at least in 1+1
dimensions) that an exactly scaling structure function can be
constructed from towers of infinitesimally narrow mesonic $q\bar q$
resonances \cite{EINHORN}.
This proof-of-principle example provides a heuristic guide to the
appearance of the qualitative features of Bloom-Gilman duality, and has
been used to motivate more elaborate studies of duality in quark models,
even though application to the baryon sector is somewhat more involved
\cite{NCBARYON}.
Given that Bloom-Gilman duality is empirically established only for
baryons (specifically, the proton), while the application of theoretical
models is generally more straightforward in the meson sector, a natural
question to consider is whether, and how, duality manifests itself
phenomenologically for the simplest $q\bar q$ system in QCD --- the pion.

As the lightest $q\bar q$ bound state, the pion plays a special role in
QCD.
Indeed, in Nature the pion presents itself as somewhat of a dichotomy:
on the one hand, its anomalously small mass suggests that it should be
identified with the pseudo-Goldstone mode of dynamical breaking of chiral
symmetry in QCD; on the other, it ought to be described equally well from
the QCD Lagrangian in terms of current quarks, with particularly
attractive forces acting in the $J^P=0^-$ channel.
The complementarity of these pictures may also reflect, loosely-speaking,
a kind of duality between the effective, hadronic description based on
symmetries, and a microscopic description in terms of partons.
This duality is effectively exploited in calculations of hadron properties
via the QCD sum rule method \cite{QCDSR}, in which results obtained in
terms of hadronic variables using dispersion relations are matched with
those of the OPE using free quarks.

In this paper we connect a number of these themes in an attempt to
further develop and elucidate the issue of quark-hadron duality for the
pion, focusing in particular on insights that can be gained from 
phenomenological constraints.
Specifically, we shall examine the possible connections between the
structure of the pion as revealed in exclusive scattering, and that
which is measured in inclusive reactions.
The latter can in principle be reconstructed given sufficient knowledge
of the form factors which parameterize transitions from the ground state
pion to excited states. 
We do not attempt this rather challenging task directly; instead, we use
the tools of the OPE to organize moments of the pion structure function
according to (matrix elements of) local operators of a given twist.
This exercise is possible because the structure function of the pion has
been determined from Drell-Yan $\pi N$ scattering data at high $Q^2$.
Of course, the absence of fixed pion targets means that the structure of
the pion at low excitation mass $W$ is not known, with the exception of
the elastic pion contribution, which has been accurately measured for
$Q^2 \alt 2$~GeV$^2$ in $\pi^+$ electroproduction off the proton.

To complement the dearth of data on specific $\pi \to \pi^*$ transitions
above threshold (but below the deep inelastic continuum), we consider a
simple model for the pion structure function in which the low $W$ spectrum
is dominated by the elastic and $\pi\to\rho$ transitions, on top of a
continuum which is estimated by evolving the leading twist structure
function to lower $Q^2$.
The discussion at low $W$ is necessarily more qualitative than for the
corresponding case of the nucleon \cite{JF2} where ample data exist.
However, even within the current limitations, this analysis provides
an estimate of the possible size of higher twist effects in the pion
structure function, and the role of the resonance region in deep
inelastic scattering (DIS) from the pion.
Preliminary results for the higher twist corrections have been presented
in Ref.~\cite{PILET}.
Here we shall extend that analysis by considering the extent to which
local duality may be valid in the pion structure function, and possible
constraints on the $x \to 1$ behavior which can be inferred from the
elastic channels.

This study is timely in view of experiments on the pion elastic
\cite{MACK,WHITE} and transition \cite{KOSSOV,BURKERT} form factors
being planned or analyzed at Jefferson Lab, which will probe the
interplay between soft and hard scattering from the pion and the onset
of perturbative QCD (pQCD) behavior.
Furthermore, recent measurements of the inclusive pion structure function
via the semi-inclusive charge-exchange reaction, $e p \to e n X$, at HERA 
have yielded some unexpected results at low $x$ \cite{HERA}, and new
experiments over a large range of $x$ are being planned at Jefferson Lab
at lower $Q^2$ \cite{KRISHNI12}.
This paper discusses the possible interelations between these
measurements, in the quest for obtaining a consistent, unified
description of the structure of the pion in electromagnetic scattering.

The structure of this paper is as follows.
After briefly reviewing in Section~II the definitions and kinematics of
inclusive lepton scattering from the pion, in Section~III we begin the
discussion by focusing on the special case of elastic scattering.
We construct an efficient parameterization of the elastic pion form
factor in the space-like region consistent with the $Q^2 \to 0$ and
$Q^2 \to \infty$ constraints.
An analysis of moments of the pion structure function is presented in
Section~IV, including the extraction of higher twists and a discussion
of the role of the resonance region.
Some of these results appeared in Ref.~\cite{PILET}.
In addition, we carefully examine the large $x$ region, which is
important for high moments, and compare predictions of several models
for the leading and higher twist contributions to the pion structure
function as $x \to 1$.
The relation of the structure function at $x \sim 1$ with the
$Q^2 \to \infty$ dependence of elastic form factors is discussed in
Section~V, where we test the hypothesis of local Bloom-Gilman duality
between the scaling structure function and the exclusive elastic and
$\pi\to\rho$ transition contributions.
Concluding remarks and a survey of future avenues for developments of
the issues presented are outlined in Section~VI.

\section{Definitions}

Inclusive scattering of an electron, or any charged lepton, from a pion,
$e \pi \to e X$, is described by the pion hadronic tensor,
\begin{eqnarray}
W_{\mu\nu}^\pi &=&
(2\pi)^3 \delta^4(p + q - p_X)
\sum_X \langle \pi | J_\mu(0) | X   \rangle
       \langle X   | J_\nu(0) | \pi \rangle\ ,
\end{eqnarray}
where $p$ and $q$ and the pion and virtual photon four-momenta,
respectively, and $p_X$ is the momentum of the hadronic final state with
invariant mass squared $W^2 = m_\pi^2 - q^2 + 2 m_\pi \nu$, with $\nu$
the energy transfer in the pion rest frame and $m_\pi$ the pion mass.
The hadronic tensor can be parameterized in terms of two structure
functions,
\begin{eqnarray}
W_{\mu\nu}^\pi
&=& \left( -g_{\mu\nu} + { q_\mu q_\nu \over q^2 } \right)\
    W_1^\pi(\nu,q^2)\
 +\ \left( p_\mu - { p \cdot q \over q^2 } q_\mu \right)
    \left( p_\nu - { p \cdot q \over q^2 } q_\nu \right)\
    { W_2^\pi(\nu,q^2) \over m_\pi^2 }\ ,
\end{eqnarray}
where $W_1^\pi$ and $W_2^\pi$ are in general functions of two variables,
for instance $\nu$ and $q^2$.
In the limit as $\nu \to \infty$ and $Q^2 \equiv -q^2 \to \infty$, with
$x = Q^2/2 p\cdot q = Q^2 / (W^2 - m_\pi^2 + Q^2)$ fixed, the functions
$W_1^\pi$ and $\nu W_2^\pi$ become scale-invariant functions of $x$,
\begin{subequations}
\begin{eqnarray}
m_\pi W_1^\pi(\nu,q^2) &\to& F_1^\pi(x)\ ,	\\
\nu W_2^\pi(\nu,q^2) &\to& F_2^\pi(x)\ .
\end{eqnarray}
\end{subequations}%
Furthermore, in this limit these functions satisfy the Callan-Gross
relation, $F_2^\pi = 2 x F_1^\pi$ \cite{CG}.
Radiative QCD corrections introduce explicit dependence of $F_{1,2}^\pi$
on the strong coupling constant, $\alpha_s(Q^2)$.
While only transversely polarized photons contribute to the $F_1^\pi$
structure function, $F_1^\pi \propto \sigma_T$, the $F_2^\pi$ structure
function receives both transverse and longitudinal contributions,
$F_2^\pi \propto \sigma_T + \sigma_L$, where $\sigma_T$ and $\sigma_L$
are the transverse and longitudinal photoabsorption cross sections,
respectively.

\section{Pion Form Factor}

The inclusive spectrum begins with the elastic peak at $W=m_\pi$,
or equivalently, $x=1$.
Because the pion is spinless, elastic scattering from the pion
contributes only to the longitudinal cross section, so that the
elastic contribution to the $F_1^\pi$ structure function vanishes.
The elastic contribution to the $F_2^\pi$ structure function is
proportional to the square of the elastic pion form factor, $F_\pi(Q^2)$,
\begin{eqnarray}
F_2^{\pi (\rm el)}(x=1,Q^2) &=&
2 m_\pi \nu\ \left( F_\pi(Q^2) \right)^2\ \delta(W^2 - m_\pi^2)
\end{eqnarray}
where $F_\pi(Q^2)$ is the elastic pion form factor.
As the most basic observable characterizing the composite nature of the
lightest bound state in QCD, the elastic form factor of the pion is of
fundamental importance to our understanding of hadronic structure.
In the approximation that the pion wave function is dominated by its
lowest $q\bar q$ Fock state, the pion form factor becomes amenable to
rigorous QCD analysis.
Indeed, it is well known that the asymptotic behavior of the pion
form factor is calculable in pQCD \cite{FJPI,LB,DM},
\begin{eqnarray}
F_\pi(Q^2) &\to&
{ 8 \pi \alpha_s(Q^2)\ f_\pi^2 \over Q^2 }
\ \ \ {\rm as}\ \ Q^2 \to \infty\ ,
\label{FpipQCD}
\end{eqnarray}
where $f_\pi = 132$~MeV is the pion decay constant.
Current data on $F_\pi$, summarized in Fig.~1, indicate that there are
large soft contributions still at $Q^2 \alt 2$~GeV$^2$
\cite{NESTERENKO,IL}.
The low $Q^2$ data are obtained from scattering 
pions off atomic electrons \cite{FF_ATOMIC}, while the higher $Q^2$ data
are taken from $^1{\rm H}(e,e'\pi^+)n$ measurements at CEA/Cornell
\cite{FF_CORNELL}, DESY \cite{FF_DESY} and JLab \cite{FF_JLAB}.
For comparison, the leading order pQCD prediction from Eq.~(\ref{FpipQCD})
is shown in Fig.~1.
Although the region of applicability of the pQCD result is
{\em a priori} unknown, the pion represents the best hope of observing the
onset of asymptotic behavior experimentally (the corresponding pQCD
calculation of the nucleon form factors significantly underestimates the
data at the same $Q^2$).

\begin{figure}[t]	
\begin{center}
\epsfysize=9cm
\leavevmode
\epsfbox{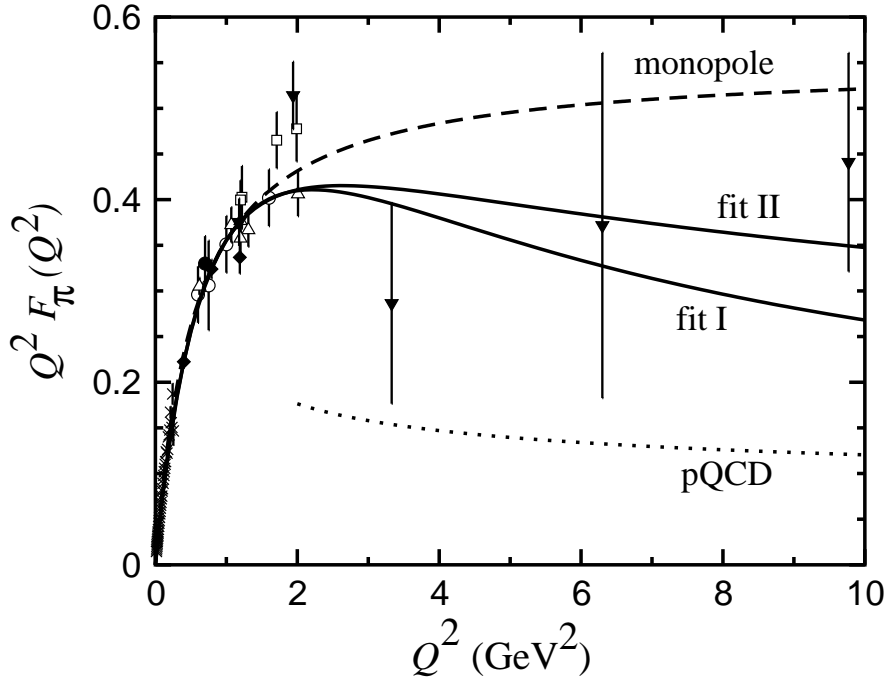}
\vspace*{0.5cm}
\caption{Pion form factor as a function of $Q^2$.
	Shown are the best fits (solid) using Eq.~(\protect\ref{Fpifit}),
	a monopole fit (dashed) with a cut-off mass of 0.74~GeV, and the
	asymptotic prediction from pQCD (dotted).}
\end{center}
\end{figure}

There have been a number of calculations of the elastic pion form
factor at low $Q^2$, for instance using the QCD sum rule approach
\cite{SMILGA}.
Rather than rely on any specific model, however, in this analysis we use
empirical data to calculate the elastic contribution to the pion structure
function.
For convenience, and for later use in Sections~IV and V, we present a
simple parameterization of the pion elastic form factor
data in the space-like region,
which is valid over the entire range of $Q^2$ currently accessible, and
smoothly interpolates between the pQCD and photoproduction limits.
For the latter, the pion form factor at low $Q^2$ can be well described
in the vector meson dominance hypothesis, in which
$F_\pi(Q^2) \sim 1/(1 + Q^2/m_\rho^2)$.
A best fit to the low $Q^2$ data using the simple monopole form is shown
in Fig.~1 (dashed), with a cut-off mass $\approx 0.74$~GeV.
The monopole fit is not compatible, however, with the behavior at high
$Q^2$ expected from pQCD.
Building in the $Q^2 \to 0$ and $Q^2 \to \infty$ constraints,
Eq.~(\ref{FpipQCD}), the available form factor data can be fitted by the
form
\begin{eqnarray}
F_\pi(Q^2) &=&
{ 1 \over 1 + Q^2/m_\rho^2 }
\left( { 1 + c_1 Z + c_2 Z^2 \over 1 + c_3 Z + c_4 Z^2 + c_5 Z^3 }
\right)\ ,
\label{Fpifit}
\end{eqnarray}
where $Z = \log(1+Q^2/\Lambda^2)$, and $\Lambda$ is the QCD scale
parameter.
The form (\ref{Fpifit}) is similar to that proposed in Ref.~\cite{WIN}
within a dispersion relation analysis, however, the form there uses 2
additional parameters, and takes a rather large value of
$\Lambda \sim 1$~GeV.
Note that the parameterization (\ref{Fpifit}) is valid only in the
space-like region; for a recent discussion of the properties of
$F_\pi(Q^2)$ in the time-like region see Ref.~\cite{GESHK}.

\begin{table}
 \begin{center}
\begin{tabular}{|l|cccc|}		\hline\hline
                &  $c_1$   &  $c_2$ &  $c_3$   &  $c_4$
					\\ \hline
\ \ fit I \ \	&$-0.201$  &  0.020 &$-0.030$  &$-0.093\ \ $
					\\
\ \ fit II\ \	&\ \ 0.100 &  0.060 &    0.538 &$-0.249\ \ $
					\\ \hline\hline
\end{tabular}
\vspace*{0.5cm}
\caption{Fit parameters for the pion form factor in
	Eq.~(\protect\ref{Fpifit}), as discussed in the text.}
 \end{center}
\end{table}

The best fit parameters $c_{1 \cdots 4}$ which give the minimum $\chi^2$
are given in Table~I.
The parameter $c_5$ is constrained by the pQCD asymptotic limit,
$c_5 = m_\rho^2 (\beta_0/32\pi^2 f_\pi^2) c_2$, where
$\beta_0 = 11 - 2 N_f/3$ (= 9 for the 3-flavor case) \cite{DONOGHUE}.
For the QCD scale parameter we take $\Lambda = 0.25$~GeV.
For completeness, we offer two parameterizations, which approach the
pQCD limit (\ref{FpipQCD}) differently: in fit~I the form factor becomes
dominated by hard scattering at $Q^2 \sim 100$~GeV$^2$, consistent with
semi-phenomenological expectations \cite{IL}, while in fit~II around half
of the strength of the form factor at this scale still comes from soft
contributions.
These are indicated by the solid lines in Fig.~1.
Better quality data are needed to constrain $F_\pi(Q^2)$ at higher $Q^2$
($\agt 2$~GeV$^2$).
To this end, there are plans to measure the pion form factor at an
energy-upgraded Jefferson Lab to $Q^2 \approx 6$~GeV$^2$ \cite{WHITE}.
%

\section{Pion Structure Function}

Going from elastic pion scattering ($W=m_\pi$) to the more general case
of inelastic scattering ($W > m_\pi$), in this section we analyze the
pion structure function, $F_2^\pi$, in terms of an OPE of its moments in
QCD, and obtain an estimate for the size of higher twists corrections to
the scaling contribution.
Following this we discuss the role of higher twists in the pion structure
function at large $x$, and compare several models for the $x \to 1$
behavior of $F_2^\pi$ with data from Drell-Yan experiments.

\subsection{Moments}

From the operator product expansion in QCD, moments of the pion
$F_2^\pi$ structure function, defined as
\begin{eqnarray}
M_n(Q^2) &=& \int_0^1 dx\ x^{n-2}\ F_2^\pi(x,Q^2)\ ,
\label{MnDEF}
\end{eqnarray}
can be expanded perturbatively at large $Q^2$ as a power series in
$1/Q^2$, with coefficients given by matrix elements of local operators
of a given twist (defined as the mass dimension minus the spin of the
operator),
\begin{eqnarray}
M_n(Q^2) &=& \sum_{k=0}^\infty {\cal A}_k^n(\alpha_s(Q^2))
	\left( { 1 \over Q^2 } \right)^k\ .
\label{MnOPE}
\end{eqnarray}
Here the leading twist (twist 2) term ${\cal A}_0^n$ corresponds to free
quark scattering, and modulo perturbative $\alpha_s(Q^2)$ corrections
is responsible for the scaling of the structure functions.
The higher twist contributions ${\cal A}_{k>0}^n$ represent matrix
elements of operators involving both quark and gluon fields, and are
suppressed by additional powers of $1/Q^2$.
The higher twist terms reflect the strength of nonperturbative QCD
effects, such as multi-parton correlations, which are associated with
confinement.

Note that the definition of $M_n(Q^2)$ includes the elastic contribution
at $x = Q^2 / (W^2 - m_\pi^2 + Q^2) = 1$, where $W$ is the mass of the
hadronic final state.
Although negligible at high $Q^2$, the elastic contribution has been
found to be important numerically at intermediate $Q^2$ for moments of
the nucleon structure function \cite{JF2}.
In the definition (\ref{MnDEF}) we use the Cornwall-Norton moments rather
than the Nachtmann moments, which are expressed in terms of the Nachtmann
scaling variable, $\xi = 2x/(1 + \sqrt{1+4x^2 m_\pi^2/Q^2})$, that
includes effects of the target mass.
The use of the Cornwall-Norton moments was advocated in Ref.~\cite{JF2}
on the grounds that it avoids the unphysical region $\xi > \xi(x=1)$.
Because of the small value of $m_\pi$, the difference between the
variables $x$ and $\xi$, and therefore between the $x$- and
$\xi$-moments, is negligible for the pion.

The seminal analysis of De~R\'ujula {\em et al.} \cite{RUJ} (see also
Ref.~\cite{JF2}) demonstrated that the onset of quark-hadron duality
is governed directly by the size of the higher twist matrix elements.
In particular, duality implies the existence of a region in the
($n$, $Q^2$) space in which the moments of the structure function are
dominated by low mass resonances, and where the higher twist
contributions are neither dominant nor negligible.
For the case of the proton $F_2$ structure function, even though there
are large contributions from the resonance region, conventionally
defined as $W \alt 2$~GeV, to the $n=2$ moment ($\sim 70\%$ at
$Q^2=1$~GeV$^2$), the higher twists contribute only around 10-20\% to
the cross section at the same $Q^2$ \cite{JF2}.
The question we wish to address here is whether there exists an
analogous region for the pion, where the resonance contributions are
important, but higher twist effects are small enough for duality to be
observed.

Of course, the distinction between the resonance region and the deep
inelastic continuum is in practice somewhat arbitrary.
In the large $N_c$ limit of QCD, for instance, the final state in DIS
from the pion is populated by infinitely narrow resonances even in
the Bjorken limit, while the structure function calculated at the quark
level produces a smooth, scaling function \cite{IJMV}.
Empirically, the spectrum of the excited states of the pion is
expected to be rather smooth sufficiently above the $\rho$ mass,
for $W \agt 1$~GeV.
Resonance excitation of heavier mesons are not expected to be easily
discernible from the DIS continuum --- the $a_1$ meson, for instance,
at a mass $W \sim 1.3$~GeV, has a rather broad width
($\sim 350-500$~MeV) \cite{PDG}.

Moments of the pion structure function can also be calculated directly
via lattice QCD, and first simulations of the leading as well as some
specific higher twist contributions have been performed \cite{LATPI}.
Although the detailed $x$ dependence, especially at large $x$ (see next
section) requires knowledge of high moments \cite{XDEP}, considerable
information on the shape of the valence distribution can already be
extracted from just the lowest 3 or 4 moments \cite{XPI}.
Calculations of a further 2 or 3 moments may be sufficient to allow both
the valence and sea distributions to be extracted as a function of $x$.

Measurements of the pion structure function have been made using the
Drell-Yan process \cite{BNLDY,FNALDY,NA3,NA10,E615} in $\pi N$
scattering, covering a large range of $x$, $0.2 \alt x \alt 1$,
and for $Q^2$ typically $\agt 20$~GeV$^2$.
It has also been extracted from the semi-inclusive DIS data at HERA for
very low $x$ and high $W$ \cite{HERA} (see also Ref.~\cite{PIONPOLE}).
However, there are no data on $F_2^\pi$ at low $W$, in the region
where mesonic resonances would dominate the cross section.
The spectrum could in principle be reconstructed by observing low $t$
neutrons produced in the semi-inclusive charge-exchange reaction,
$e p \to e n X$, where $t$ is the momentum transfer squared between
the proton and neutron, and extrapolating to the pion pole to ensure
$\pi$ exchange dominance.
In the meantime, to obtain a quantitative estimate of the importance of
the resonance region, we model the pion spectrum at low $W$ in terms of
the elastic and $\rho$ pole contributions, on top of the DIS continuum
evolved down from the higher $Q^2$ region, as outlined in
Ref.~\cite{PILET}.
The leading twist structure function can be reconstructed from
parameterizations \cite{GRVPI,SMRSPI,GRSPI} of quark distributions in
the pion obtained from global analyses of the pion Drell-Yan data.
Unless otherwise stated, in this work we use the low $Q^2$ fit from
Ref.~\cite{GRVPI}, which gives the leading twist parton distributions
in the pion for $Q^2 > 0.25$~GeV$^2$
(our conclusions are not sensitive to the use of other parameterizations
\cite{SMRSPI,GRSPI}).
For the elastic contribution we use the parameterization in
Eq.~(\ref{Fpifit}) [fit~I].

The contribution of the $\rho$ meson is described by the $\pi\to\rho$
transition form factor, $F_{\pi\rho}(Q^2)$, which is normalized such
that $F_{\pi\rho}(0)=1$, and is expected to fall as $1/Q^4$ at large
$Q^2$ (compared with $1/Q^2$ for $F_\pi(Q^2)$).
Since there is no empirical information on $F_{\pi\rho}(Q^2)$, we
consider several models in the literature, based on a relativistic
Bethe-Salpeter vertex function \cite{ITOGROSS}, a covariant
Dyson-Schwinger approach \cite{MARIS}, and light-cone QCD sum rules
\cite{KHODJ}.
These represent a sizable range ($\sim 100\%$) in the magnitude of
$F_{\pi\rho}(Q^2)$ over the region of $Q^2$ covered in this analysis,
with the calculation of Ref.~\cite{KHODJ} giving a somewhat smaller
result than those in Refs.~\cite{ITOGROSS,MARIS}.
The spread in these predictions can be viewed as an indicator of the
uncertainty in this contribution.
The $\pi\to\rho$ transition form factor can be extracted, for instance,
from $\rho$ electroproduction data off the proton, $e p \to e p \rho^0$
\cite{KOSSOV,BURKERT}, by reconstructing the decay of the $\rho^0$ into
two pions.
It also forms an important input into the calculation of meson-exchange
current contributions to deuteron form factors at large $Q^2$
\cite{DEUTFF}.

\begin{figure}[t]	
\begin{center}
\epsfig{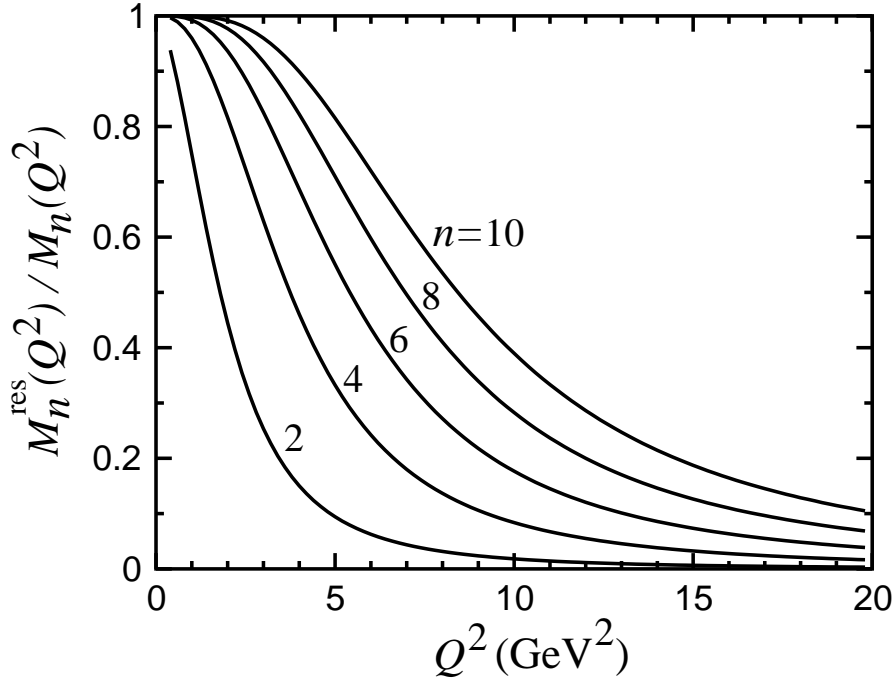}
\vspace*{0.5cm}
\caption{Contributions to moments of the pion structure function from
	the resonance region, $W < W_{\rm res} = 1$~GeV, relative to the
	total \protect\cite{PILET}.}
\end{center}
\end{figure}

The contributions from the ``resonance region'',
$W < W_{\rm res} \equiv 1$~GeV, to the moments of the pion structure
function,
\begin{eqnarray}
M_n^{\rm res}(Q^2) &=& \int_{x_{\rm res}}^1 dx\ x^{n-2} F_2^\pi(x,Q^2)\ ,
\label{MnRES}
\end{eqnarray}
are plotted in Fig.~2 as a ratio to the total moment,
for $n = 2, \cdots, 10$.
The integration in $M_n^{\rm res}(Q^2)$ is from
$x_{\rm res} = Q^2/(W_{\rm res}^2 - m_\pi^2 + Q^2)$
to the elastic point, $x=1$.
The low $W$ region contributes as much as 50\% at $Q^2 = 2$~GeV$^2$
to the total $n=2$ moment, decreasing to $\alt 1\%$ for
$Q^2 \agt 10$~GeV$^2$ \cite{PILET}.
Higher moments are more sensitive to the large $x$ region, and
subsequently receive larger contributions from low $W$.
The $n=10$ moment, for example, is almost completely saturated by the
resonance region at $Q^2 = 2$~GeV$^2$, and even at $Q^2 = 10$~GeV$^2$
still receives some 40\% of its strength from $W < 1$~GeV even at
$Q^2 = 10$~GeV$^2$.

\begin{figure}[t]	
\begin{center}
\epsfig{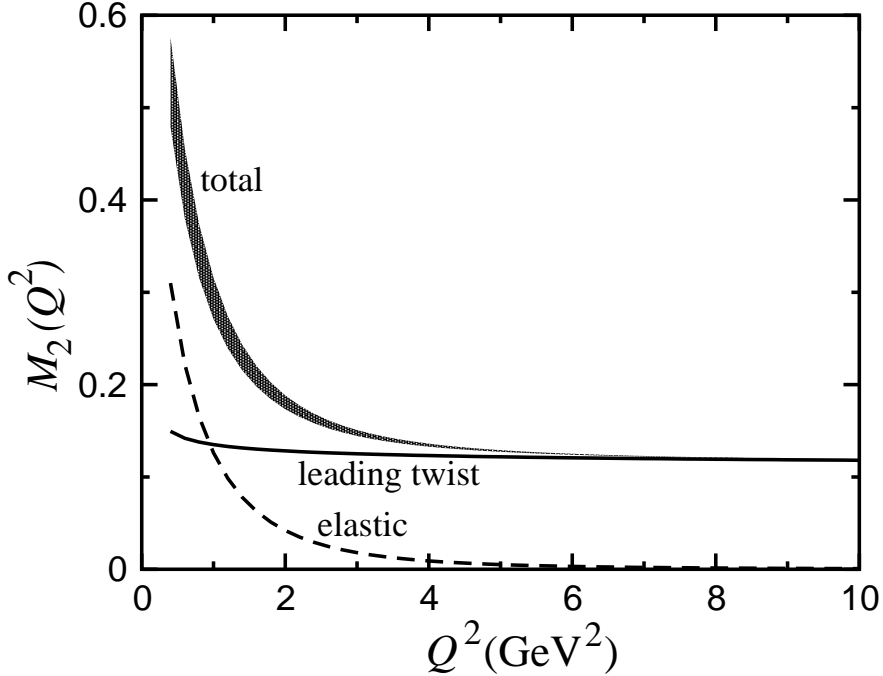}
\vspace*{0.5cm}
\caption{Lowest ($n=2$) moment of the pion structure function.
	The leading twist (solid) and elastic (dashed)
	contributions are shown, and the shaded region represents
	the total moment using different models for the $\pi\to\rho$
	transition \protect\cite{PILET}.}
\end{center}
\end{figure}

The relatively large magnitude of the resonance contributions suggests
that higher twist effects play a more important role in the moments of
the pion structure function than for the case of the nucleon.
In Fig.~3 the lowest ($n=2$) moment of $F_2^\pi$ is displayed, together
with its various contributions.
The leading twist component,
\begin{eqnarray}
M_n^{\rm LT}(Q^2) &=& \int_0^1 dx\ x^{n-2} F_{2, {\rm LT}}^\pi(x,Q^2)\ ,
\label{MnLT}
\end{eqnarray}
is expressed (at leading order in $\alpha_s(Q^2)$) in terms of the
twist-2 quark distributions in the pion,
\begin{eqnarray}
F_{2, {\rm LT}}^\pi(x,Q^2) &=& \sum_q e_q^2\ x q^\pi(x,Q^2)\ ,
\label{F2piLT}
\end{eqnarray}
where the valence part of $q^\pi$ is normalized such that
$\int dx\ q^\pi_{\rm val}(x,Q^2) = 1$.
The leading twist contribution is dominant at $Q^2 > 5$~GeV$^2$, while
the deviation of the total moment from the leading twist at lower $Q^2$
indicates the increasingly important role played by higher twists there.
While negligible beyond $Q^2 \approx 4$~GeV$^2$, the elastic
contribution is as large as the leading twist already at
$Q^2 \approx 1$~GeV$^2$.
The $\pi\to\rho$ contribution is more uncertain, and the band in Fig.~3
represents the total moment calculated using different models
\cite{ITOGROSS,MARIS,KHODJ} of $F_{\pi\rho}(Q^2)$.
However, while the current uncertainty in this contribution is
conservatively taken to be $\sim 100\%$, doubling this would lead to a
modest increase of the band in Fig.~3.
%
%
Uncertainty from poor knowledge of the leading twist distributions at
small $x$ \cite{GRVPI,SMRSPI,GRSPI} is not expected to be large.

\begin{figure}[t]	
\begin{center}
\epsfig{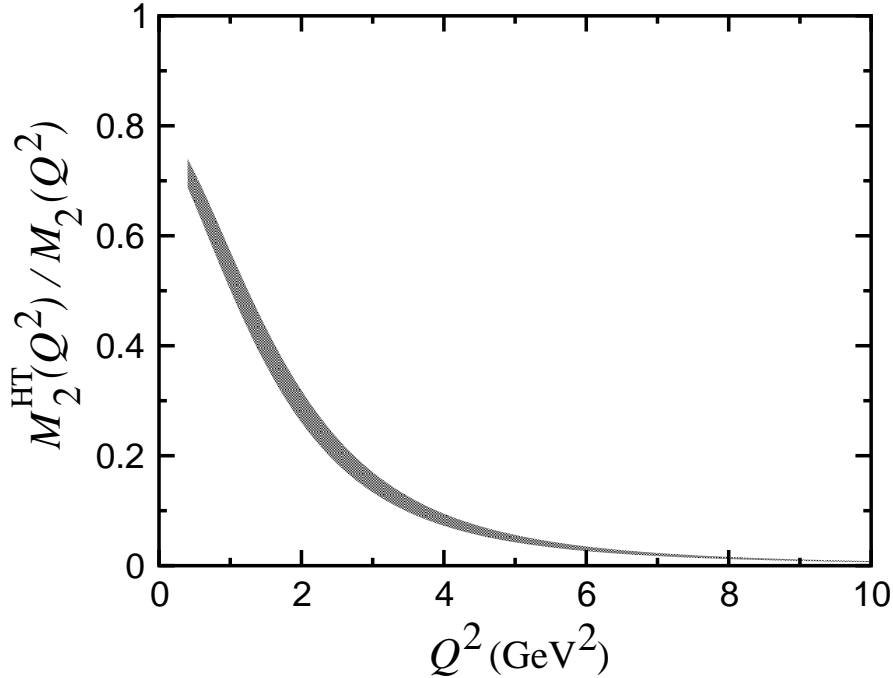}
\vspace*{0.5cm}
\caption{Higher twist contribution to the $n=2$ moment of the pion
	structure function, as a ratio to the total moment.
	The band indicates the uncertainty due to the model dependence
	of the $\pi\to\rho$ transition form factor \protect\cite{PILET}.}
\end{center}
\end{figure}

To extract the higher twist part of the moments, one needs to subtract
the leading twist contribution in Eq.~(\ref{MnLT}) from the total moments
$M_n(Q^2)$,
\begin{eqnarray}
M_n^{\rm HT}(Q^2)
&=& M_n(Q^2) - M_n^{\rm LT}(Q^2) - M_n^{\rm TM}(Q^2)\ ,
\end{eqnarray}
where $M_n^{\rm TM}(Q^2)$ arises from target mass corrections.
Because the target mass correction, which is formally of leading twist,
is proportional to $m_\pi^2/Q^2$, its contribution will only be felt
when $Q^2 \sim m_\pi^2$, which is far from the region where the twist
expansion is expected to be valid.
In principle nonperturbative effects can mix higher twist with higher
order effects in $\alpha_s$, rendering the formal separation of the two
problematic \cite{MUELLERHT,RENORMALON,NNNLO,ALEKHIN}.
Indeed, the perturbative expansion itself may not even be convergent.
However, by restricting the kinematics to the region of $Q^2$ in which
the $1/Q^2$ term is significantly larger than the next order correction
in $\alpha_s$, the ambiguity in defining the higher
twist terms can be neglected \cite{JF2}.
In Fig.~4 the higher twist contribution to the $n=2$ moment is displayed
as a function of $Q^2$.
The band again represents an estimate of the uncertainty in the
$\pi\to\rho$ transition form factor, as in Fig.~3.
At $Q^2 = 1$~GeV$^2$ the higher twist contribution is as large as the
leading twist, decreasing to $\sim 1/3$ at $Q^2 = 2$~GeV$^2$, and
vanishes rapidly for $Q^2 \agt 5$~GeV$^2$.

As observed in Ref.~\cite{PILET}, the size of the higher twist
contribution at $Q^2 \sim 1$~GeV$^2$ appears larger than that found in
similar analyses of the proton $F_2$ \cite{JF2} and $g_1$ \cite{JG1}
structure functions.
This can be qualitatively understood in terms of the intrinsic
transverse momentum of quarks in the hadron, $\langle k_T^2 \rangle$,
which typically sets the scale of the higher twist effects.
%
%
Since the transverse momentum is roughly given by the inverse size of the
hadron, $\langle k_T^2 \rangle \sim 1/R^2$, the smaller confinement radius
of the pion means that the average $\langle k_T^2 \rangle$ of quarks in 
the pion will be larger than that in the nucleon.
Therefore the magnitude of higher twists in $F_2^\pi$ is expected to be
somewhat larger (${\cal O}(50\%)$) than in $F_2^p$.
The E615 Collaboration indeed finds the value
$\langle k_T^2 \rangle = 0.8 \pm 0.3$~GeV$^2$,
within the higher twist model of Ref.~\cite{BB}.
%
The experimental value is obtained by analyzing the $x \to 1$ dependence
of the measured $\mu^+\mu^-$ pairs produced in $\pi N$ collisions, and
the angular distribution at large $x$.
We discuss this in more detail below.

\subsection{$x \to 1$ Behavior}

The $x \to 1$ behavior of structure functions is important for several
reasons.
As discussed in the previous section, higher moments of $F_2^\pi$ receive
increasingly large contributions from the large-$x$ region, so that a
reliable extraction of higher twists from data requires an accurate
determination of quark distributions at $x \sim 1$.
In addition, since the $x \sim 1$ region is dominated by the lowest
$q\bar q$ Fock state component of the pion light-cone wave function, in
which the interacting quark carries most of the momentum of the pion,
the behavior of the structure function at $x \to 1$ is expected to be
correlated with that of the elastic form factor at $Q^2 \to \infty$.
In this section we review various predictions for $F_2^\pi$ in the
limit as $x \to 1$, and relate these to the effects of higher twists
discussed in the previous section on the $x \to 1$ behavior of the
structure function.

Working within a field theoretic parton model framework which predates
QCD, Drell and Yan \cite{DY} and West \cite{WEST} showed that if the
asymptotic behavior of the form factor is $(1/Q^2)^n$, then the structure
function should behave as $(1-x)^{2n-1}$ as $x \to 1$.
This is referred to as the Drell-Yan-West (DYW) relation.
Simple application to the case of the pion, in which the elastic form
factor behaves as $1/Q^2$ at large $Q^2$, leads to the prediction
\begin{eqnarray}
F_2^\pi(x\to 1) &\sim& (1-x)\ .
\end{eqnarray}
This behavior is also predicted in the model of Ref.~\cite{GBB}.

A dynamical basis for the exclusive--inclusive relation was provided
with the advent of QCD.
By observing that the interacting quark at large $x$ is far off its mass
shell, Farrar and Jackson \cite{FJ} derived the $x \to 1$ behavior of the
structure function at $x \to 1$ by considering perturbative one gluon
exchange between the $q$ and $\bar q$ constituents in the lowest Fock
state component of the pion wave function.
They found a characteristic $\sim (1-x)^2$ dependence for the transverse
part of $F_2^\pi$, in apparent contradiction with the naive DYW relation
(the breakdown of the DYW relation for spinless hadrons was discussed
earlier by Landshoff and Polkinghorne \cite{LANDSHOFF}).
The longitudinal cross section was found to scale like $1/Q^2$ relative
to the transverse \cite{FJ}.
Using so-called `softened' field theory \cite{BRODFAR}, in which the
pion-quark vertex function is described by a Bethe-Salpeter type
equation, Ezawa \cite{EZAWA} found a similar $(1-x)^2$ behavior.

Gunion {\em et al.} \cite{GUNION} later generalized the gluon exchange
description to include subleading $1/Q^2$ corrections for both the
$F_2^\pi$ and longitudinal $F_L^\pi$ structure functions at $x \to 1$,
\begin{eqnarray}
F_2^\pi(x) &\sim&
S_2 (1-x)^2 + { T_2 \over Q^2}\ ,	\\
\label{F2PRED}
F_L^\pi(x) &\sim& { S_L \over Q^2 }\ ,
\label{FLPRED}
\end{eqnarray}
where the constants $S_2$, $T_2$ and $S_L$ are determined
phenomenologically.
More generally, according to the pQCD `counting rules' \cite{LB}, the
leading components for any hadron with $n$ spectator (non-interacting)
partons were found \cite{GUNION} to behave as
$(1-x)^{2n - 1 + 2|\Delta S_z|}$ in the $x \to 1$ limit,
where $\Delta S_z$ is the difference between the helicities of the
hadron and the interacting quark.
%
%
More recently, other nonperturbative models have been used to calculate
the pion structure function \cite{PIONMODELS}, however, because of
difficulties associated with incorporating high momentum components of
the wave function, these may not be reliable in the $x \sim 1$ region.

\begin{figure}[t]	
\begin{center}
\epsfig{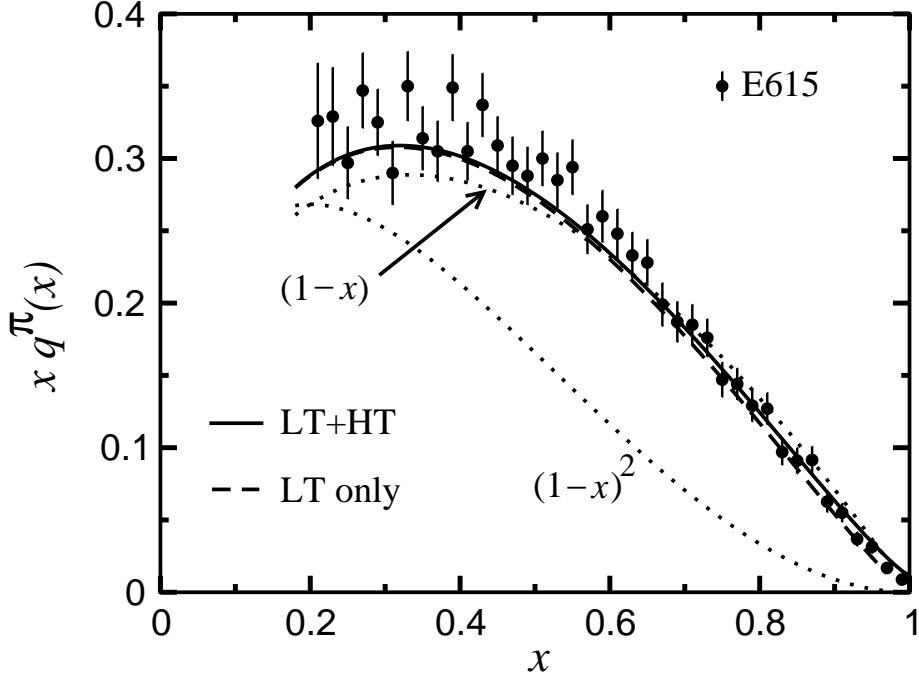}
\vspace*{0.5cm}
\caption{Valence quark distribution in the pion extracted from the FNAL
	E615 Drell-Yan experiment \protect\cite{E615}, fitted with
	leading twist (dashed) and leading + higher twist (solid)
	contributions, as in Eq.~(\protect\ref{e615fit}).
	The functional forms $(1-x)$ and $(1-x)^2$ (dotted) are shown
	for comparison.}
\end{center}
\end{figure}

The predicted $x \to 1$ behavior of the pion structure function can be
tested by comparing with Drell-Yan data.
The $x$ dependence of the pion quark distributions has been measured in
Drell-Yan $\mu^+\mu^-$ pair production in $\pi N$ collisions (in practice,
$\pi A$) at BNL \cite{BNLDY}, CERN \cite{NA3,NA10} and at Fermilab
\cite{FNALDY,E615}.
The data for $q^\pi(x) \equiv u^{\pi^+}(x) = \bar d^{\pi^+}(x)$ from the
most recent Fermilab experiment \cite{E615} are shown in Fig.~5 for
$4.05 < m_{\mu\mu} < 8.55$~GeV, where $m_{\mu\mu}$ is the invariant mass
of the $\mu^+\mu^-$ pair.
The scale dependence within this region was found to be small \cite{E615}.
The data were fitted using the form
\begin{eqnarray}
q^\pi(x,Q^2) &=&
N x^a (1-x)^b + \gamma\ { 2\ x^2 \over 9\ Q^2 }\ ,
\label{e615fit}
\end{eqnarray}
where $N$ is a constant, fixed by normalization, and the scale $Q^2$
is identified with the the dimuon mass squared, $m_{\mu\mu}^2$.
The form (\ref{e615fit}) parameterizes both leading and higher twist
effects.
Including corrections from $Q^2$ evolution, the best fit value for the
exponent governing the $x \to 1$ behavior was found to be
$b \approx 1.21$--1.30 \cite{E615}, consistent with the findings of
the earlier CERN experiments \cite{NA3,NA10}.
The result of the leading twist E615 fit with $b=1.27$ is shown in
Fig.~5 (dashed).
The forms $(1-x)$ \cite{DY,WEST,GBB} and $(1-x)^2$
\cite{EZAWA,LB,FJ,GUNION,MUELLERX1} are also shown for comparison at
large $x$ (dotted).
The data clearly favor a shape closer to $(1-x)$, rather than the
$(1-x)^2$ shape implied by the counting rules \cite{LB}.

It has been suggested \cite{BB} that higher twist effects in the pion
structure function could obscure the true leading twist behavior.
The higher twist coefficient $T_2$ was calculated in a pQCD-inspired
model by Berger and Brodsky \cite{BB} in terms of the intrinsic quark
momentum in the pion,
\begin{eqnarray}
T_2 &=& { 2 \over 9 } { \langle k_T^2 \rangle \over Q^2 }\ .
\end{eqnarray}
Since it is independent of $x$, it can be argued \cite{BB} that the
higher twist contribution may in fact dominate the scaling term at
fixed $Q^2 (1-x)$ as $Q^2 \to \infty$, and mimic the observed $(1-x)$
dependence if $\langle k_T^2 \rangle \approx 1$~GeV$^2$.
Conway {\em et al.} \cite{E615} subsequently performed an analysis of
the E615 data by fitting also the term $\gamma$ in Eq.~(\ref{e615fit}).
The extracted value of $b$ was found to be largely independent of
the value of $\gamma$ chosen.
To investigate whether the quadratic term may be masked by an additional
component not included in the model \cite{BB}, Conway {\em et al.}
searched for a nonzero intercept of $F_2^\pi$ at $x=1$.
The fit with $\gamma = 0.83$~GeV$^2$ was found to be only marginally
better than that with $\gamma=0$ (the significance being 2.5 standard
deviations), although the fit at $x \sim 1$ was also sensitive to the
input nucleon sea distributions in the analysis of the Drell-Yan data.
The leading + higher twist fit with $\gamma = 0.83$~GeV$^2$ is shown in
Fig.~5 (solid curve).
The effect on the overall fit is indeed quite marginal, although at very
large $x$ ($\agt 0.9$) the differences between this and the pure leading
twist fit are more apparent.

Mueller \cite{MUELLERX1} has pointed out that Sudakov effects, which
introduce terms like $\alpha_s(Q^2) \ln^2(1/(1-x))$ into the $x \to 1$
analysis, may invalidate the usual renormalization group analysis of
DIS at large $x$.
Including power and double logarithmic corrections, one finds that the
$x \to 1$ behavior of $F_2^\pi$ in this case becomes \cite{MUELLERX1}
\begin{eqnarray}
F_2^\pi &\sim& (1-x)^2
\exp\left\{ -{ 4 C_F \over \beta_0 }
	    \left[ \ln{1\over 1-x} \ln\ln Q^2
		 - { \ln^2(1/(1-x)) \over 2 \ln Q^2 }
		 - \ln{1\over 1-x} \ln\ln{1\over 1-x}
	    \right]
     \right\}\ .	\nonumber\\
& &
\label{Mueller}
\end{eqnarray}
When $\ln(1/(1-x)) = {\cal O}(\ln Q^2/\ln\ln Q^2)$ higher twist terms
compete with the leading twist, and the dominant contribution is then
from the longitudinal structure function, which behaves as
$F_L^\pi \sim (1/Q^2) \ln(Q^2(1-x))$.
Taking this criterion literally, for $Q^2 \sim 1$~GeV$^2$ this would
occur at $x \sim 0.93$, while for $Q^2 \sim 100$~GeV$^2$ the higher
twists would be expected to dominate at $x \agt 0.97$.
A cautionary note regarding Eq.~(\ref{Mueller}), however, is that single
logarithmic effects have not been included in the analysis, and their
effects on Eq.~(\ref{Mueller}) are unclear.
Further discussion of these effects can be found in
Refs.~\cite{MUELLERHT,MUELLERX1}.
Carlson and Mukhopadhyay \cite{CMQ2} have also studied the effects of
radiative corrections on the $x \to 1$ behavior of the structure
function, and the appearance of higher twists in the low-$W$ region.
In particular, the scale dependence of the $(1-x)$ exponent was found
to be $(1-x)^{b + c \ln\ln Q^2}$, with $c$ calculable perturbatively.
The $Q^2$ dependence of the pion structure function at $x \sim 1$
clearly deserves further study.

A cleaner signature of high twist effects at large $x$ comes from the
angular distribution of dimuon pairs produced in Drell-Yan collisions.
The angular dependence of the Drell-Yan cross section is given by
\cite{LAMTUNG} (see also Ref.~\cite{BRAND})
\begin{eqnarray}
{ d\sigma \over d\Omega }
&\propto& 1 + \lambda \cos^2\theta + \mu \sin 2\theta \cos\phi
	+ {\nu \over 2} \sin^2\theta \cos 2\phi\ ,
\end{eqnarray}
where the angles $\theta$ and $\phi$ are defined in the $\mu^+\mu^-$
rest system, 
and $\lambda$, $\mu$ and $\nu$ are functions of the kinematic variables.
%
In the model of Ref.~\cite{BB}, the leading twist $(1-x)^2$ term is
associated with a $(1+\cos^2\theta)$ dependence, while the higher twist
$\langle k_T^2 \rangle / Q^2$ term has a characteristic $\sin^2\theta$
dependence.
In particular, the transverse cross section corresponds to $\lambda=1$,
while deviation from a pure $(1+\cos^2\theta)$ dependence would indicate
the presence of longitudinal or higher twist contributions.
The data \cite{E615} are consistent with $\lambda=1$ for $x \alt 0.6$,
while the larger-$x$ data show clear deviations from pure transverse
scattering, suggesting the presence of higher twist contributions at
these $x$ values.
With the fitted value of $\beta$ ($\approx 1.2$--1.3), the measured $x$
dependence of $\lambda$ could be accommodated with
$\langle k_T^2 \rangle \approx 0.8$~GeV$^2$.
Using the value $\beta=2$ predicted by the pQCD counting rules, the
observed $\lambda$ values could be made to fit the data by requiring
that $\langle k_T^2 \rangle \sim 0.1$~GeV$^2$.
However, in addition to being much smaller than the value
$\langle k_T^2 \rangle \sim 1$~GeV$^2$ suggested in Ref.~\cite{BB},
this scenario is disfavored by a direct comparison with the $x$
dependence of $q^\pi(x)$, as discussed above.


While the values of the quark intrinsic transverse momentum extracted
from the Drell-Yan data are consistent with the size of higher twist
effects observed in Sec.~III, there does appear to be a clear conflict
between the counting rule predictions for the $x \to 1$ behavior of
$F_2^\pi$ and the empirical $x$ dependence.
Several reasons could account for this discrepancy.
Even higher twist effects, beyond those of twist-4 parameterized in
Eq.~(\ref{e615fit}), could be present and obfuscate an underlying
$(1-x)^2$ leading twist behavior.
This appears unlikely, however, given the relatively large $Q^2$ values
($Q^2 \agt 20$~GeV$^2$) at which the data are sampled, and the rapid
fall off of the higher twist contributions to the moments observed in
Sec.~IV.A.

On the other hand, as alluded to above, the extraction of the pion
structure function requires as input the parton distributions in the
nucleon.
Since the bulk of the data for $x > 0.5$ corresponds to a nucleon
light-cone momentum fraction $x_N \approx 0.05$--0.1, errors may be
introduced into the analysis through poor knowledge of the sea quark,
or (at higher order) gluon, distributions in the nucleon.
Furthermore, because the data are taken on nuclear targets (e.g. tungsten
for the E615 experiment), nuclear effects may give rise to corrections to
the nucleon quark distributions, especially in the region $x_N \sim 0.05$,
where nuclear shadowing is known to play an important role \cite{SHAD}.
The effects of using more modern nucleon parton distributions, and
including nuclear corrections in the analysis, are currently being
investigated \cite{KRISHNI}.

It may also be that the asymptotic behavior does not set in until $x$ is
very close to 1, and that the functional form (\ref{e615fit}) is simply
too restrictive to adequately reflect this behavior, in which case a
more sophisticated parameterization
would be required.
Further, nonobservation of the predicted counting rule behavior may not
necessarily imply a breakdown of pQCD.
The derivation of the counting rules for large-$x$ structure functions
from Feynman diagrams in terms of hard gluon exchanges between quarks
involves an infrared cut-off mass parameter, $m$, which regulates the
integrals when $k_T \to 0$ \cite{GUNION,HOODBHOY}.
Although an analysis based on pQCD should be valid also for $m=0$, the
counting rule results are sensitive to the parameter, $m$, and comparison
with phenomenology requires a nonzero value \cite{GUNION}.

Regardless of the ultimate $x \to 1$ behavior of $F_2^\pi$ extracted
from data, it is instructive to examine whether the asymptotic
inclusive--exclusive relations between the pion structure function and
the pion elastic and transition form factors at large $Q^2$ can provide
additional constraints.
In the next section we use local quark-hadron duality to study these
relations in more detail.

\section{Local Quark-Hadron Duality}

There has been a revival of interest recently in the phenomenon of
Bloom-Gilman duality in electron--nucleon scattering.
This has been stimulated largely by recent high precision measurements
\cite{NICU} at Jefferson~Lab of the $F_2$ structure function of the
proton, which demonstrated that duality works remarkably well for each
of the prominent low-lying resonance regions, including the elastic
\cite{EL,MEL}, as well as for the integrated structure function,
to rather low values of $Q^2$.
Ongoing and planned future studies will focus on duality in other
structure functions, such as $g_1$ \cite{FOREST} and $F_L$ \cite{ERIC},
and for hadrons other than the proton.

While the existence of local quark-hadron duality appears inevitable in
QCD at asymptotically large momenta \cite{IJMV,DUALMOD}, it is not
{\em a priori} clear that it should work at finite $Q^2$.
Indeed, there are reasons why at low $Q^2$ it should not work at all
\cite{IJMV}, and its appearance may in principle be due to accidental
cancellations (due to quark charges in the proton \cite{CI,GOTT}, for
instance) of possibly large higher twist effects.
A systematic study of local duality for other hadrons, such as the pion,
is therefore crucial to revealing the true origin of this phenomenon.

Shortly after the original observations of Bloom-Gilman duality for the
proton \cite{BG}, generalizations to the case of the pion were explored.
By extending the finite-energy sum rules \cite{FESR} devised for the
proton duality studies, Moffat and Snell derived a local duality sum rule
relating the elastic pion form factor with the scaling structure function
of the pion \cite{MOFFAT},
\begin{eqnarray}
[F_\pi(Q^2)]^2
&\approx& \int_1^{\omega_{\rm max}} d\omega\ \nu W_2^\pi(\omega)\ ,
\label{local}
\end{eqnarray}
where $\nu W_2^\pi \equiv F_2^\pi$ is a function of the scaling
variable $\omega = 1/x$.
The upper limit $\omega_{\rm max} = 1 + (W^2_{\rm max}-m_\pi^2)/Q^2$
was set in Ref.~\cite{MOFFAT} by $W_{\rm max} \approx 1.3$~GeV, in order
for the integration region to include most of the effect of the hadron
pole, and not too much contribution from higher resonances \cite{MOFFAT}.
To test the validity of the finite-energy sum rule relation (\ref{local}),
Moffat and Snell \cite{MOFFAT}, and later Mahapatra \cite{MAHAPATRA},
constructed Regge-based models of the pion structure function (their
analyses predated the Drell-Yan pion structure function measurements
\cite{BNLDY,FNALDY,NA3,NA10,E615}) to compare with the then available
pion form factor data.

The existence of Drell-Yan data on $F_2^\pi$ now allows one to test this
relation quantitatively using {\em only} phenomenological input.
Using parameterizations of the $F_2^\pi(x)$ data from Ref.~\cite{E615}
(see Fig.~5), the resulting form factor $F_\pi(Q^2)$ extracted from
Eq.~(\ref{local}) is shown in Fig.~6.
The agreement appears remarkably good.
On the other hand, the magnitude of the form factor depends somewhat on
the precise value chosen for $W_{\rm max}$, so the agreement in Fig.~6
should not be taken too literally.
Nevertheless, the shape of the form factor is determined by the $x$
dependence of the structure function at large $x$.
In particular, while a $(1-x)$ behavior leads to a similar $Q^2$
dependence to that for the E615 fit,
assuming a $(1-x)^2$ behavior gives a form factor which drops more
rapidly with $Q^2$.
This simply reflects the kinematic constraint $(1-1/\omega) \sim 1/Q^2$
at fixed $W$.

\begin{figure}[t]	
\begin{center}
\epsfig{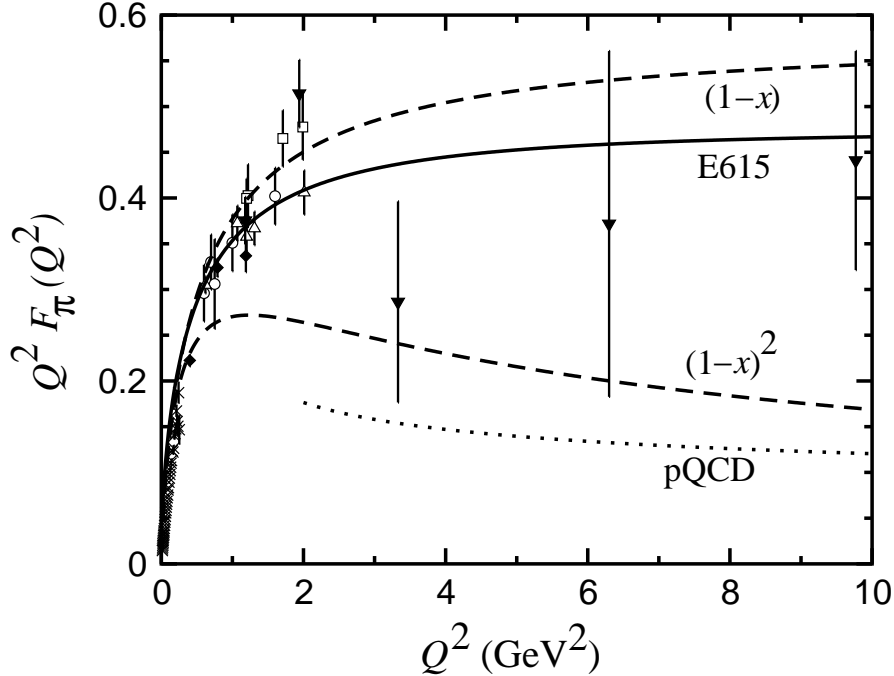}
\vspace*{0.5cm}
\caption{Local duality prediction for the pion form factor, using
	phenomenological pion structure function input from the FNAL
	E615 Drell-Yan experiment \protect\cite{E615} (solid), and the
	forms $F_2^\pi(x) \sim (1-x)$ and $(1-x)^2$ (dashed).
	The asymptotic leading order pQCD prediction (dotted) is shown
	for reference.}
\end{center}
\end{figure}

Although the apparent phenomenological success of the local duality
relation (\ref{local}) is alluring, there are theoretical reasons why
its foundations may be questioned.
In fact, the workings of local duality for the pion are even more
intriguing than for the nucleon.
Because it has spin 0, elastic scattering from the pion contributes only
to the longitudinal cross section ($F_T^\pi(x=1,Q^2)=0$).
On the other hand, the spin 1/2 nature of quarks guarantees that the deep
inelastic structure function of the pion is dominated at large $Q^2$ by
the transverse cross section \cite{RUJ,FJ}.
Taken at face value, the relation (\ref{local}) would suggest a nontrivial
duality relation between longitudinal and transverse cross sections.
Whether local duality holds individually for longitudinal and transverse
cross sections, or for their sum, is currently being investigated
experimentally.
Indications from proton data are that indeed some sort of duality holds
for both the transverse and longitudinal structure functions of the
proton individually \cite{NICU,ERIC}.

While the elastic form factor of the pion is purely longitudinal, the
$\pi \to \rho$ transition is purely transverse.
It has been suggested \cite{RUJ} that the average of the pion elastic
and $\pi\to\rho$ transition form factors may instead dual the deep
inelastic pion structure function at $x \sim 1$.
If we take the simple model used in Sec.~IV for the low $W$ part of
the pion structure function, in which the inclusive pion spectrum at
$W \alt 1$~GeV is dominated by the $\pi \to \pi$ and $\pi \to \rho$
transitions, we can estimate the degree to which such a duality may
be valid.
Generalizing Eq.~(\ref{local}) to include the lowest-lying longitudinal
and transverse contributions to the structure function, one can replace
the left hand side of (\ref{local}) with
$[F_\pi(Q^2)]^2 + \omega_\rho [F_{\pi\rho}(Q^2)]^2$, where
$\omega_\rho = 1 + (m_\rho^2 - m_\pi^2)/Q^2$.

The sum of the lowest two `resonance' contributions (elastic + $\rho$)
to the generalized finite-energy sum rule is shown in Fig.~7 as a ratio
to the corresponding leading twist DIS structure function over a similar
range of $W$.
The upper and lower sets of curves envelop different models
\cite{ITOGROSS,MARIS,KHODJ} of $F_{\pi\rho}(Q^2)$, which can be seen as
an indicator of the current uncertainty in the calculation.
Integrating to $W_{\rm max} = 1$~GeV, the resonance/DIS ratio at
$Q^2 \sim 2$~GeV$^2$ is $\sim 50 \pm 30\%$ above unity, and is consistent
with unity for $Q^2 \sim 4$--6~GeV$^2$ (solid curves).
As a test of the sensitivity of the results to the value of $W_{\rm max}$,
the resonance/DIS ratio is also shown for $W_{\rm max} = 1.3$~GeV
(dotted curves).
In this case the agreement is better for $Q^2 \sim 1$--3~GeV$^2$,
with the ratio being $\sim 30 \pm 20\%$ below unity for
$Q^2 \sim 4$--6~GeV$^2$.

\begin{figure}[t]	
\begin{center}
\epsfig{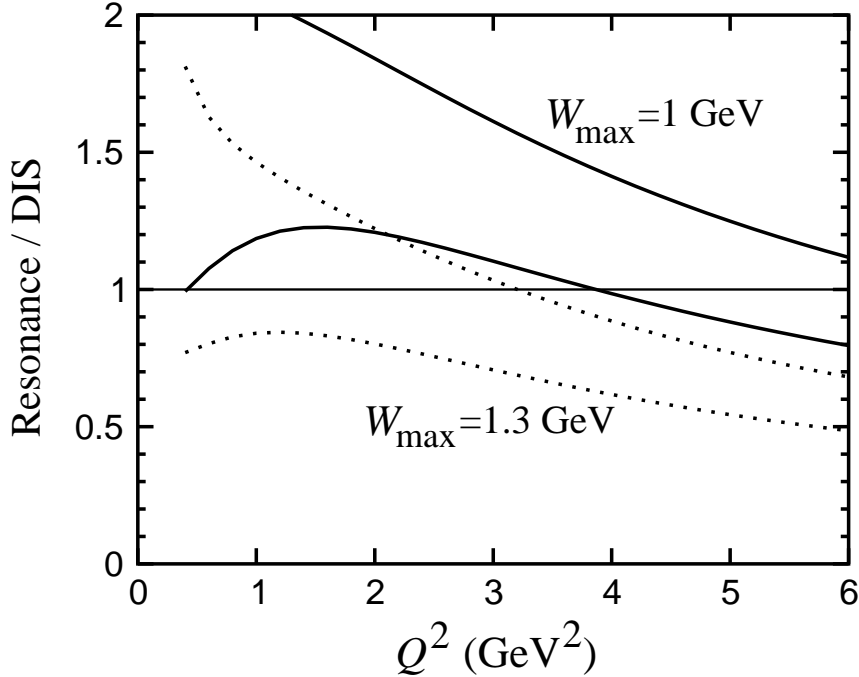}
\vspace*{0.5cm}
\caption{Ratio of the pion resonance (elastic + $\pi\to\rho$ transition)
	contributions relative to the DIS continuum, for different values
	of $W_{\rm max}$.  The two sets of upper and lower curves
	reflect the uncertainties in the $\pi\to\rho$ transition form
	factor.}
\end{center}
\end{figure}

Given the simple nature of the model used for the excitation spectrum,
and the poor knowledge of the $\pi\to\rho$ transition form factor, as
well as of the pion elastic form factor beyond $Q^2 \approx 2$~GeV$^2$,
the comparison can only be viewed as qualitative.
However, the agreement between the DIS and resonance contributions
appears promising.
Clearly, data on the inclusive $\pi$ spectrum at low $W$ would be
invaluable for testing the local duality hypothesis more quantitatively.
In addition, measurement of the individual transverse and longitudinal
cross sections of the pion, using Rosenbluth separation techniques,
would allow duality to be tested separately for the longitudinal and
transverse structure functions of the pion.

\section{Conclusion}

Understanding the structure of the pion represents a fundamental
challenge in QCD.
High energy scattering experiments reveal its quark and gluon
substructure, while at low energies its role as a Goldstone boson mode
associated with chiral symmetry breaking in QCD is essential in
describing the long-range structure and interactions of hadrons.
We have sought to elucidate the structure of the pion by considering
its response to electromagnetic probes, focusing in particular on the
connection between inclusive and exclusive channels.

The relation between the pion structure function and the pion elastic
and transition form factors has been studied in the context of
quark-hadron duality.
Moments of the pion structure function have been evaluated, and the
role of the resonance region studied, assuming that the low $W$
resonant spectrum is dominated by the elastic and $\pi\to\rho$
transitions.
The contribution of the resonance region ($W \alt 1$~GeV) to the lowest
moment of $F_2^\pi$ is $\sim 50\%$ at $Q^2 \approx 2$~GeV$^2$, and only
falls below 10\% for $Q^2 \agt 5$~GeV$^2$.
The elastic component, while negligible for $Q^2 \agt 3$~GeV$^2$, is
comparable to the leading twist contribution at $Q^2 \approx 1$~GeV$^2$.
Combined, this means that the higher twist corrections to the $n=2$
moment are $\sim 50\%$ at $Q^2 = 1$~GeV$^2$, $\sim 30\%$ at
$Q^2 = 2$~GeV$^2$, and only become insignificant beyond
$Q^2 \approx 6$~GeV$^2$.

Uncertainties on these estimates are mainly due to the poor knowledge of
the inclusive pion spectrum at low $W$, which limits the extent to which
duality in the pion can be tested quantitatively.
Only the elastic form factor has been accurately measured to
$Q^2 \approx 2$~GeV$^2$, although at larger $Q^2$ it is poorly
constrained.
The inclusive pion spectrum can be extracted from data from the
semi-inclusive charge-exchange reaction, $e p \to e n X$, at low $t$,
for instance at Jefferson Lab \cite{BURKERT}.
This could also allow one to determine the individual exclusive channels
at low $W$.
In addition, a Rosenbluth separation would allow the transverse and
longitudinal structure functions to be extracted.
%

Within the current uncertainties, the higher twist effects in the pion
appear larger than the analogous corrections extracted from moments of
the nucleon structure functions \cite{JF2,JG1}.
This can be generically understood in terms of the larger intrinsic 
transverse momentum of quarks, which governs the scale of the $1/Q^2$
corrections, in the pion than in nucleon, associated with the smaller
pion confinement radius.
The implication is that duality would therefore be expected to set in
later (at larger $Q^2$) for the pion than for the nucleon.

Higher twist effects have also been observed in the pion structure
function at large $x$ by the E615 Collaboration at Fermilab \cite{E615}.
The $x$ dependence and angular distribution of $\mu^+\mu^-$ pairs
produced in $\pi N$ collisions at $x \sim 1$ suggests a value
$\langle k_T^2 \rangle = 0.8 \pm 0.3$~GeV$^2$, which is larger than the
typical quark transverse momentum in the nucleon (${\cal O}$(500~MeV)).
On the other hand, the measured $x$ dependence appears to be harder than
that predicted by counting rules \cite{LB} or models based on
perturbative one gluon exchange \cite{EZAWA,BB,FJ,GUNION}, favoring a
$(1-x)$ shape over a $(1-x)^2$ dependence.
A reanalysis \cite{KRISHNI} of the Drell-Yan data to take into account
nuclear corrections and updated sea quark distributions in the nucleon,
which are used as input into the analysis, is necessary for a definitive
assessment of the validity of the various approaches.
Additional modification of the $x \to 1$ behavior due to Sudakov-like
effects \cite{MUELLERX1} may also need to be considered before drawing
final conclusions about the implications of the observed $x \to 1$
dependence.
There are also plans to measure $F_2^\pi$ in semi-inclusive reactions
over a range of $x$ at Jefferson Lab \cite{KRISHNI12} to confirm the
Drell-Yan and semi-inclusive HERA measurements, which should allow a
more thorough exploration of the higher twist effects at lower $Q^2$.

The specific $x \to 1$ behavior of the pion structure function has
consequences for the $Q^2$ dependence of the elastic pion form factor,
if one assumes the validity of local quark-hadron duality for the pion.
In particular, using parameterizations of the Drell-Yan structure
function data, the existing data on $F_\pi(Q^2)$ can be fitted if the
upper limit of the integration region above the elastic peak extends to
$W_{\rm max} \approx 1.3$~GeV.
Analogous fits with a $(1-x)^2$ shape fall off too rapidly with $Q^2$
and do not fit $F_\pi(Q^2)$ as well.

On the other hand, there may be limitations of the extent to which local
duality can hold for the pion, as such duality implies a nontrivial
relationship between the longitudinal and transverse cross sections.
It may in fact be more appropriate to examine whether the sum of the
longitudinal (elastic) and transverse ($\pi\to\rho$ transition)
contributions duals the DIS structure function at low $W$.
Using phenomenological models for the $\pi\to\rho$ form factor, our
estimate for the sum of the lowest-lying resonant contributions is in
qualitative agreement with the corresponding scaling contribution in the
same $W$ interval.
However, empirical information on the strength and $Q^2$ dependence of
$F_{\pi\rho}(Q^2)$ is necessary for a more quantitative test.
The $\pi\to\rho$ transition form factor can in practice be extracted
from $\rho$ electroproduction data \cite{KOSSOV}.
At larger $Q^2$, the $\pi\to\rho$ transition is expected to be suppressed
relative to the elastic contribution, and to test the local duality here
will require a more accurate determination of $F_\pi(Q^2)$.
The pion form factor $F_\pi(Q^2)$ will soon be measured to
$Q^2 = 2.5$~GeV$^2$ at Jefferson Lab \cite{MACK} in $\pi^+$
electroproduction from the proton, and possibly to $Q^2 = 6$~GeV$^2$
with an energy upgraded facility \cite{WHITE}.

Finally, this analysis can be easily extended to the strangeness sector,
to study the duality between the form factor and structure function of
the kaon.
Data from the Drell-Yan reaction in $K^-$--nucleus collisions
\cite{KAONSF} indicate that the quark distribution in the kaon is
similar to that in the pion, and measurements of the kaon form factor,
$F_K(Q^2)$, have also recently been reported \cite{KAONFF}.
Future measurements of $F_K(Q^2)$ at larger $Q^2$ ($\sim 2$~GeV$^2$)
\cite{MACK} would allow the first quantitative test of local Bloom-Gilman
duality in strange hadrons.

\acknowledgements

Helpful discussions with V.~Burkert, R.~Ent, P.~Hoodbhoy, D.~Mack,
and K.~Wijesooriya are gratefully acknowledged.
I would also like to thank F.~Gross, P.~Maris and P.~Tandy for sending
the results of their form factor calculations.
This work was supported by the U.S. Department of Energy contract
\mbox{DE-AC05-84ER40150}, under which the Southeastern Universities
Research Association (SURA) operates the Thomas Jefferson National
Accelerator Facility (Jefferson Lab).


\end{document}